\documentclass[lettersize,journal,comsoc]{IEEEtran}
\usepackage{amsmath,amsfonts}
\usepackage{algorithmic}
\usepackage{algorithm}
\usepackage{array}
\usepackage[caption=false,font=normalsize,labelfont=sf,textfont=sf]{subfig}
\usepackage{textcomp}
\usepackage{stfloats}
\usepackage{url}
\usepackage{verbatim}
\usepackage{graphicx}
\usepackage{color}
\usepackage{cite}

\begin{document}

\title{\LARGE Quantum Pulse Gate Attack on IM/DD Optical Key Distribution\\
Exploiting Symbol Shape Distortion}

\author{Marcin Jarzyna, Micha{\l} Jachura, Konrad Banaszek,~\IEEEmembership{Senior Member, IEEE}
\thanks{This work was supported by the project ``Quantum Optical
Technologies'' carried out under the International Research Agendas Programme
of the Foundation for Polish Science co-financed by the European Union through
the European Regional Development Fund. {\em (Corresponding author: Konrad
Banaszek.)}}
\thanks{Marcin Jarzyna and Micha{\l} Jachura are with the Centre for Quantum Optical Technologies, Centre of New Technologies, University of Warsaw, Banacha 2c, 02-097 Warszawa, Poland (e-mail: m.jarzyna@cent.uw.edu.pl;
m.jachura@cent.uw.edu.pl). Konrad Banaszek is with the Centre for Quantum Optical Technologies, Centre of New Technologies, University of Warsaw, Banacha 2c, 02-097 Warszawa, Poland, and also with the Faculty of Physics, University
of Warsaw, Pasteura 5, 02-093 Warszawa, Poland (e-mail: k.banaszek@uw.edu.pl).}}



\maketitle

\begin{abstract}
Intensity modulation/direct detection (IM/DD) optical key distribution (OKD) is a method to generate a secret key whose security against passive eavesdropping is guaranteed by the shot noise inherent to the photodetection process. Here the effects of intensity-dependent symbol shape distortion on the IM/DD OKD security are investigated assuming that the eavesdropper can implement temporal mode demultiplexing using e.g.\ the quantum pulse gating technique. The quantitative analysis includes key generation based on either hard- or soft-decoding of the detected signal as well as the impact of excess detection noise.
A simple rule-of-thumb relation between the severity of the symbol shape distortion and the signal strength required to ensure key security is presented.
\end{abstract}


\section{Introduction}
\IEEEPARstart{P}{resently}, a substantial effort is dedicated to fortifying optical communication systems with physical layer security solutions \cite{BlochBarros2011}. Among various approaches currently being pursued, secret random keys shared between distant nodes can be generated using techniques of quantum key distribution (QKD) \cite{PirandolaAOP2020}, or by exploiting shot noise inevitably accompanying detection of optical radiation \cite{IkutaNJP2016}. The latter technique, henceforth referred to as optical key distribution (OKD), places less stringent requirements on the physical system implementation compared to QKD while offering protection against passive, beam-splitting-type eavesdropping attacks that are conceivable using current or near-term technology. Remarkably, OKD can be implemented in intensity modulation/direct detection (IM/DD) communication systems \cite{YamamoriJJAP2020}, providing security against passive eavesdropping even if the fraction of the signal captured by an adversary (Eve) is much larger than that collected by the legitimate recipient (Bob) \cite{BanaszekOpEx2021}. This makes IM/DD OKD an attractive option to ensure physical layer security in free-space optical links that could be incorporated e.g.\ in future satellite-to-earth optical communication systems \cite{TrinhIEEETC2020,JachuraICSO2022}.

The purpose of this Letter is to analyze the security of IM/DD OKD in a practically relevant scenario when the signal intensity modulation applied by the sender (Alice) is associated with a change of the shape of optical pulses encoding distinct key bit values. Such symbol shape distortions can arise e.g.\ due to the nonlinear characteristics of the modulator used to carve out OKD pulses from a cw source laser beam. While a pulse shape modification does not constitute a security threat if Eve measures only the total optical energy of received pulses, as assumed in previous security analyses \cite{IkutaNJP2016,YamamoriJJAP2020,BanaszekOpEx2021,TrinhIEEETC2020}, the presence of such distortions opens up a possibility to deploy more powerful eavesdropping strategies based on separating the captured signal into a set of orthogonal temporal modes while retaining their individual quantum statistical properties. Such temporal demultiplexing can be implemented using the quantum pulse gate (QPG) technique based on mode-selective three-wave mixing of the optical signal with suitably shaped pump pulses \cite{BrechtReddyPRX2015}. Currently the QPG technology undergoes rather swift progress with prospective applications such as noise reduction in classical and quantum communications
\cite{BanaszekSPIE2019,RaymerBanaszekOpEx2020} and improving the resolution of time-delay measurements \cite{AnsariPRXQ2021}. It is worth noting that concurrently even more powerful time-frequency signal processing techniques are being developed \cite{MazelanikNCOMM2022}.

This paper is organized as follows. Sec.~\ref{Sec:PhysicalSetup} describes the physical system for OKD including the QPG attack by the adversary. The principle of secure key distribution is presented in Sec.~\ref{Sec:KeySecurity} and its generation rates are analyzed quantitatively in Sec.~\ref{Sec:KeyRates}. Finally, Sec.~\ref{Sec:Conclusions} concludes the paper.

\section{Physical system}
\label{Sec:PhysicalSetup}

In the binary-modulated IM/DD OKD protocol, Alice's transmitter Tx$_A$ prepares in each temporal slot a symbol in the form of a light pulse with one of two optical energies characterized by the mean photon number $n_0$ or $n_1$ corresponding to the two equiprobable key bit values $q_A=0,1$ that are chosen by Alice at random. The modulation depth is chosen sufficiently low so that the detection shot noise fundamentally prevents either Bob or Eve to identify the value of every transmitted key bit. However, as discussed in Sec.~\ref{Sec:KeySecurity},  a suitable reconciliation protocol makes it possible for Alice and Bob to postselect events that will yield a secure key unknown to Eve.

\begin{figure}[h]
\centering
\includegraphics[width=1\columnwidth]{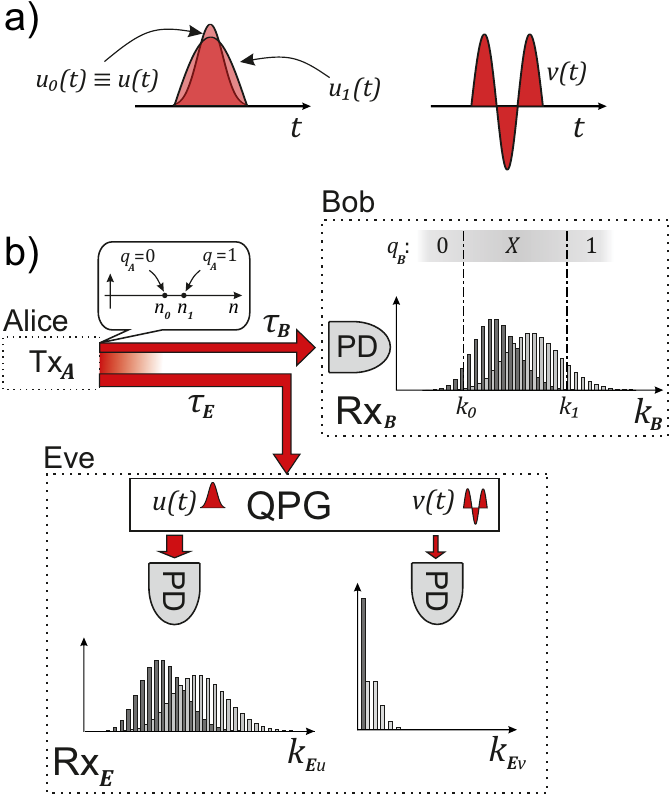}
\caption{(a) The pair of temporal modes $u_0(t)$ and $u_1(t)$ corresponding to two optical energies $n_0$ and $n_1$ generated by Alice's transmitter and the normalized mode $v(t)$ that describes the component of $u_1(t)$ orthogonal to $u_0(t)$ calculated according to Eq.~(\ref{Eq:v(t)def}). (b) The physical system under consideration. Alice's transmitter Tx$_A$ emits pulses with one of two slightly different optical energies $n_0$ or $n_1$ depending on the key bit value $q_A=0,1$ randomly chosen by Alice for each pulse.
In Bob's receiver Rx$_B$ the fraction $\tau_B$ of a pulse received by Bob produces photocounts on the detector PD whose number $k_B$ follows a Poisson distribution with a mean that depends on the key bit value chosen by Alice. In the hard-decoding scenario Bob retains outermost events assigning to them the key bit values $q_B=0$ or $q_B=1$ and the remaining events ${\mathsf X}$ are removed from further processing which is communicated to Alice over a public channel. The fraction $\tau_E$ of the signal received by Eve is temporally demultiplexed using a quantum pulse gate QPG into modes $u(t) \equiv u_0(t)$ and the orthogonal complement $v(t)$ defined in Eq.~(\ref{Eq:v(t)def}), followed by photon counting. For ideal shot-noise limited photodetection, registering one or more photocounts in the channel $v(t)$ unambiguously indicates a pulse emitted in the mode $u_1(t)$ corresponding to Alice's key bit value $q_A=1$.}
\label{Fig:Setup}
\end{figure}

In order to account for the possibility of symbol shape distortion, the two optical energies $n_0$ and $n_1$ will be associated with two complex temporal pulse envelopes $u_0(t)$ and $u_1(t)$ that are normalized to one, $\int dt \, |u_0(t)|^2 =\int dt \, |u_1(t)|^2 =1$, as illustrated in Fig.~\ref{Fig:Setup}(a). The symbol shape distortion will be characterized using the parameter
\begin{equation}
\label{Eq:Distortion}
{\cal D} = 1-\left| \int dt \, u^\ast_0(t) u_1(t) \right|^2,
\end{equation}
that is equal to zero when the two pulse envelopes are identical,  and approaches one for orthogonal envelope functions. The parameter ${\cal D}$ represents the fraction of the optical power that is lost when one filters out a given temporal mode from the signal prepared in the other mode, in full analogy with spatial mode filtering in optical waveguides \cite{Haus1987}.

As shown in Fig.~\ref{Fig:Setup}(b)  the legitimate recipient, Bob, receives a fraction $\tau_B$ of the signal power sent by Alice. His receiver Rx$_B$ measures the light intensity with a photon counting detector PD. In order to keep the notation concise, the transmission factor $\tau_B$ is taken to include also Bob's detector efficiency. The photocount number $k_B$ obtained from Bob's measurement is characterized by a conditional Poisson distribution
\begin{equation}
\label{Eq:k_B|q_A}
k_B|q_A \sim \textrm{Pois}(\tau_B n_{q_A}+n_b), \qquad q_A =0,1.
\end{equation}
The above expression accounts for excess background noise by adding to the distribution mean the parameter $n_b$ that can include both stray radiation collected by Bob as well as detector dark counts.

The pulse shape distortion associated with intensity modulation opens up for Eve the possibility to implement the following attack shown in Fig.~\ref{Fig:Setup}(b).
The signal captured by Eve, whose relative power constitutes a fraction $\tau_E$ of the signal transmitted by Alice, is separated in the receiver Rx$_E$ into two orthogonal temporal modes characterized by complex envelopes
$u(t) \equiv u_0(t)$ and
\begin{equation}
\label{Eq:v(t)def}
v(t) \equiv \frac{1}{\sqrt{\cal D}}\left( u_1(t) - u_0(t) \int dt' \, u_0^\ast(t')u_1(t')\right)
\end{equation}
that specifies the part of $u_1(t)$ which is orthogonal to $u_0(t)$. The factor $1/\sqrt{\cal D}$ ensures normalization of $v(t)$. The three modes involved in the problem are depicted schematically in Fig.~\ref{Fig:Setup}(a). Separation into orthogonal temporal modes, even if overlapping in the time domain, can be realized using the QPG technique \cite{BrechtReddyPRX2015}.

In the next stage of Eve's receiver Rx$_E$, light carried by the modes $u(t)$ and $v(t)$ is detected individually with photon counting detectors as shown in Fig.~\ref{Fig:Setup}(b).
It will be convenient to denote the two pulse optical energies received by Eve corresponding to Alice's bit values $q_A=0,1$ as $n_{E0} = \tau_E n_0$ and $n_{E1} = \tau_E n_1$ respectively.
Depending on Alice's chosen key bit value $q_A$
the statistics of photocount numbers $k_{Eu}$ and $k_{Ev}$ on detectors monitoring demultiplexed modes $u(t)$ and $v(t)$ are given by:
\begin{align*}
q_A & = 0: & k_{Eu} & \sim \textrm{Pois}(n_{E0}), & k_{Ev} & \sim \textrm{Pois}(0), \\
q_A & = 1: & k_{Eu} & \sim \textrm{Pois}((1-{\cal D}) n_{E1}), & k_{Ev} &
\sim \textrm{Pois}({\cal D}n_{E1}).
\end{align*}
The pair of photocount numbers $k_u, k_v$ constitutes the information available to Eve to learn about the key that is generated between Alice and Bob. Operation of Eve's receiver Rx$_E$ with shot-noise limited detection at 100\%\ efficiency has been assumed here. The model described above corresponds to the worst-case eavesdropping scenario where Eve possesses full knowledge about symbol shapes obtained. e.g. from access to the transmitter design or thorough prior characterization of the signal generated by Alice. The results presented below can be interpreted as lower bounds on the attainable key rates when Eve's knowledge about symbol shape distortion is incomplete.


\section{Secure key}
\label{Sec:KeySecurity}

\begin{figure}[h]
\centering
\includegraphics[width=0.8\columnwidth]{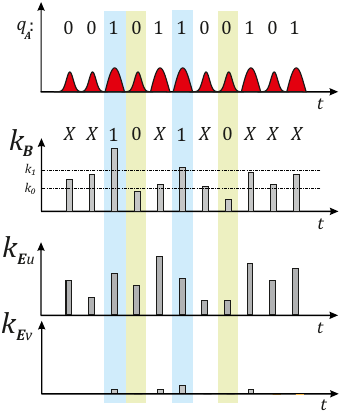}
\caption{Generation of the secure key. When Bob selects only outermost values of the photocount number $k_B < k_0$ or $k_B > k_1$, the bit values for retained events will be nearly perfectly correlated with the bit values $q_A$ chosen by Alice. Without pulse shape distortion, Eve's knowledge about the generated key is severely limited as for a given pulse optical energy the photocount number $k_{Eu}$ is statistically independent of $k_{B}$. However, when pulse shape distortion results in redirecting some of the optical energy to the photodetector monitoring the mode $v(t)$, the photocount number $k_{Ev} >0$ unambiguously indicates the bit value $q_A=1$.}
\label{Fig:KeyGeneration}
\end{figure}

The simplest and the most intuitive method for Alice and Bob to generate the cryptographic key is to apply hard decoding to the photocount number $k_B$ detected by Bob by setting two thresholds $k_0$ and $k_1$ and using the following discrimination recipe:
\begin{equation}
q_B = \begin{cases} 0, & \text{if $k_B < k_0 $,} \\
{\sf X}, & \text{if $ k_0 \le k_B \le k_1$,} \\
1, & \text{if $ k_B > k_1$}
\end{cases}
\label{Eq:qBdef}
\end{equation}
that is shown schematically in Fig.~\ref{Fig:KeyGeneration}.

Inconclusive outcomes ${\sf X}$ are communicated to Alice over a public channel and removed from further processing. For thresholds $k_0$ and $k_1$ set sufficiently far apart in the outermost regions of the photocount statistics the outcomes $q_B=0,1$ will be nearly perfectly correlated with the key bit values $q_A$ chosen by Alice. Residual errors can be removed by implementing an error correction protocol \cite{LiverisIEEECL2002,ElkoussITWIT2010}. Eve's knowledge about the key is limited by the fact that for a given optical energy $n_0$ or $n_1$ of the pulse transmitted by Alice the photocount numbers registered by Bob's and Eve's detectors are statistically uncorrelated. Without pulse shape distortion, when $u_0(t)=u_1(t)$, the outermost events that are postselected by Bob to generate the key will correspond at Eve's receiver to randomly distributed outcomes $k_{Eu}|q_A \sim \textrm{Pois}(\tau_E n_{q_A})$ and hence on average will carry less information about the key bit value compared to that available to Bob. This observation underlies the security of the generated key, which can be refined by means of privacy amplification to remove completely any remaining Eve's knowledge. In the presence of symbol shape distortion, Eve's ability to detect a signal demultiplexed in the set of orthogonal temporal modes described in Sec.~\ref{Sec:PhysicalSetup} opens up a potentially more powerful eavesdropping strategy. Namely, it is straightforward to see in Fig.~\ref{Fig:KeyGeneration} that detection of one or more photons in the mode $v(t)$ unambiguously identifies Alice's bit value as $q_A=1$, thus revealing to Eve much more information about the key.

The quantitative analysis of the key security in the presence of symbol shape distortion will be based on the Csisz\'{a}r-K\"{o}rner expression for the attainable key per slot in the reverse reconciliation scenario which reads \cite{CsiszarKornerIEEETIT1978}
\begin{equation}
\label{Eq:Key}
{\mathsf K} = \max \{ {\mathsf I}(A;B) - {\mathsf I}(B;E) , 0\}.
\end{equation}
Here ${\mathsf I}$ is the mutual information, the label $A$ stands for Alice's binary variable $q_A=0,1$, whereas $B$ corresponds to Bob's detection outcome and $E$ includes both Eve's variables $k_{Eu}$ and $k_{Ev}$. It is assumed here that key reconciliation can be implemented with 100\%\ information theoretic efficiency. Error correcting codes approaching such efficiency have been described in the context of QKD implementations \cite{ElkoussITWIT2010, ElkoussQIC2011,Johnson2015}. Two decoding scenarios will be considered in the following. For the soft-decoding scenario $B$ stands for the actual photocount number $k_B$, whereas for the hard-decoding scenario it is the discriminated variable $q_B$ defined in Eq.~(\ref{Eq:qBdef}).

\section{Key rates}
\label{Sec:KeyRates}

\begin{figure*}[h]
\centering
\includegraphics[width=1.73\columnwidth]{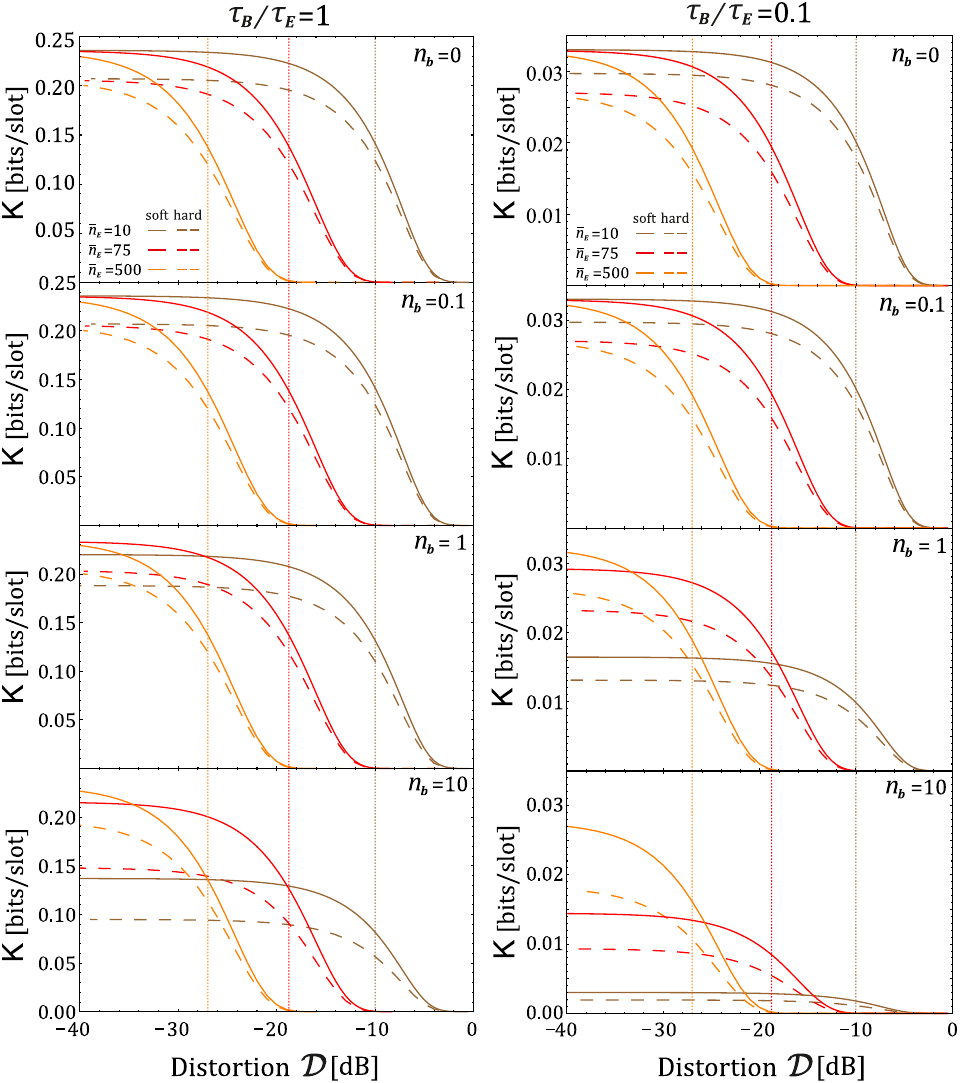}
\caption{Attainable secure key rates in bits per slot as a function of the pulse shape distortion ${\cal D}$ in decibels [dB] defined in Eq.~(\ref{Eq:Distortion}) for the ratio of transmission factors to Bob's and Eve's receivers equal to $\tau_B/\tau_E =1$ (left column) and $\tau_B/\tau_E =0.1$ (right column).
Three signal strengths characterized by the average pulse optical energy received by Eve
$\bar{n}_E = 10$ (brown), $\bar{n}_E = 75$ (red), and $\bar{n}_E = 500$ (orange) have been used assuming soft (solid lines) and hard (dashed lines) decoding of the signal detected by Bob. The two panels in the top row depict the case with no excess noise at Bob's receiver, $n_b=0$, with increasing noise amount $n_b=0,0.1,1,$ and $10$ shown in subsequent rows underneath.}
\label{Fig:Plots}
\end{figure*}

The attainable secure key rate ${\sf K}$ defined in Eq.~(\ref{Eq:Key}) will be analyzed as a function of the distortion parameter
${\cal D}$ defined in Eq.~(\ref{Eq:Distortion}) and the signal strength. The latter can be conveniently characterized with the average pulse optical energy detected by Eve $\bar{n}_E = (n_{E0} + n_{E1})/2$. With this choice of parameters, the performance of Bob's receiver is determined by the transmission ratio $\tau_B/\tau_E$ and the background photocount number $n_b$.
For given values of these parameters the key rate needs to be optimized with respect to the signal modulation depth that can be characterized using a rescaled parameter
\begin{equation}
\delta_E = (n_{E1} - n_{E0})/ (2\sqrt{\bar{n}_E})
\end{equation}
and, in the hard decoding scenario, additionally over the discrimination thresholds  used in Eq.~(\ref{Eq:qBdef}). The results are shown in Fig.~\ref{Fig:Plots} for the ratio $\tau_B/\tau_E=1$ (left column), when Bob and Eve collect the same fraction of the signal, and for $\tau_B/\tau_E=0.1$ (right column), when Eve has the capacity to collect ten times as much signal compared to Bob, using e.g. a telescope with a larger aperture in a free-space optical communication scenario. It is worth noting that typical key rates shown in Fig.~\ref{Fig:Plots} are substantially higher than those reported in \cite{Micius} for a prepare-and-measure QKD demonstration between a low Earth orbit satellite and a ground station, which amount to approx.\ $3\times 10^{-6}$ secure bit per slot. This is because of more stringent security assumptions in the latter case, including Eve's ability to access and manipulate in an arbitrary manner the optical signal at any stage after leaving Alice's transmitter.

 The graphs in Fig.~\ref{Fig:Plots} depict the attainable key per slot as a function of the pulse shape distortion ${\cal D}$ in decibels [dB] for different signal strengths $\bar{n}_E$ (coded with colors) and increasing amount of the excess noise $n_b$ (top to bottom). Starting with the soft-decoding case, shown in Fig.~\ref{Fig:Plots} with solid lines, several observations are in place. In the absence of background noise at Bob's receiver, $n_b=0$, shown in the top panels of Fig.~\ref{Fig:Plots}, the key rate tends to the same value for vanishing symbol shape distortion, ${\cal D} \rightarrow 0$ regardless of the signal strength. This can be related to the fact that the key security is a consequence of photocount fluctuations rather than the absolute signal strength.  However, the signal strength starts to have an important role when the shape distortion comes into play: the higher the signal strength $\bar{n}_E$, the lower the value of the distortion parameter for which Eve starts to gain substantial information about the key. This can be related to the observation made in Sec.~\ref{Sec:KeySecurity} that detecting just one photocount in the mode $v(t)$ is sufficient to identify the key bit value chosen by Alice as $q_A=1$. Indeed, the values ${\cal D} = 1/\bar{n}_E$, shown with vertical lines in Fig.~\ref{Fig:Plots} indicate quite well when the symbol shape distortion starts to reduce substantially the attainable key rate.
Thus, a rule-of-thumb condition for the symbol shape distortion to have a negligible effect on the key security is ${\cal D}\bar{n}_E \ll 1$.

The impact of excess background noise $n_b$ in Bob's receiver on the attainable key can be analyzed by inspecting panels below the top row in Fig.~\ref{Fig:Plots}. It is seen that for negligible pulse shape distortion, a higher signal strength reduces the effects of background noise, as its relative contribution to the photocount statistics becomes less significant. However, at the same time a higher signal strength implies a stronger susceptibility to pulse shape distortion, as discussed in the preceding paragraph for the case $n_b=0$. For a given level of pulse shape distortion and the amount of background noise one can identify an optimal signal strength that gives the maximum key value by counterbalancing the two effects discussed above. Finally, one can see in Fig.~\ref{Fig:Plots} that hard decoding of the signal detected by Bob with optimized discrimination thresholds lowers the key by a relatively small amount compared to the soft-decoding case, while it may substantially simplify the error correcting step.

\section{Conclusions}
\label{Sec:Conclusions}

The simplicity of IM/DD optical communication systems makes them an attractive option as the physical layer to generate a cryptographic key using the OKD technique which ensures security against passive eavesdropping. Security analysis of implementations of key distribution protocols needs to include so-called side channel attacks that are facilitated by a richer physical structure of the optical carrier of information compared to that assumed in the protocol principle of operation. In the case of the OKD protocol, modification of the pulse shape associated with the symbol value has been shown to enable eavesdropping based on temporal mode demultiplexing followed by photon counting. One possible remedy to this threat is to keep the strength of the signal captured by an eavesdropper at a sufficiently low level so that the optical energy carried by individual demultiplexed modes effectively does not allow for identification of the key bit value.
Alternatively, a strategy to suppress the pulse shape distortion would be to generate the OKD signal by modulating, instead of continuous wave laser source, a train of pulses substantially shorter than the temporal slot duration. This would avoid the effects of transient modulator transmission. In this setting, as long as the modulator transmission is flat over the pulse window, the symbol shape is determined solely by the input pulse waveform, independently of the modulated optical energy. On a final note, one should mention that correlations between the symbol value and the modal---temporal or spectral---structure of the emitted electromagnetic field are an issue also in standard QKD protocols, both in discrete-variable \cite{BiswasIEEEJQE2021}
as well as continuous-variable variants \cite{JainQST2021}.


\section*{Acknowledgment}
Insightful discussions with B. Brecht, P. Kolenderski, M. Lasota, and C. Silberhorn are gratefully acknowledged.

\bibliography{imddokd}

\bibliographystyle{myIEEEtran}

\end{document}